\documentstyle[12pt,epsf]{article}
\catcode`\@=11 \@addtoreset{equation}{section}  
\renewcommand{\theequation}                     
         {\arabic{section}.\arabic{equation}}   
\title{Nonlinear modification of quantum mechanics}
\author{P. Leifer}
\date{Mortimer and Raymond Sackler Institute of Advanced Studies \\ Tel-Aviv University, Tel-Aviv 69978, Israel \\
e-mail:leifer@ccsg.tau.ac.il}
\begin{document}
\maketitle
\begin{abstract}
In order to prevent ``unavoidable'' break-down of the 
``peaceful coexistence'' between foundations of quantum theory and relativity I propose a new type of a quantum gauge theory (superrelativity). This differs from 
ordinary gauge theories in the sense that the affine connection of this theory is constructed from first derivatives of the Fubini-Study metric tensor in the projective Hilbert space of the pure quantum states CP(N). That is we have not merely analogy with general relativity but this construction should presumably provide a unification of general relativity and quantum theory.
\end{abstract} 


\section{Introduction}
The interesting project of the generalized nonlinear 
quantum mechanics of Weinberg \cite{W1} was 
subjected to the critical analysis \cite{Gis} in 
the framework of probabilistic interpretation with respect to a possibility of the ``faster-than-light communication''
in nonlinear quantum dynamics. But now it has beeb shown \cite{Gold} that a specific nonlinear scheme may be successfull in spite of this general criticism. 

The general inspiration for my approach to the nonlinear 
version of quantum theory is as followes:
Laws of non-relativistic classical physics are indifferent
to {\it any type of forces}. But development of the theory of such {\it natural type of force} as electromagnetic demands the relativistic generalization of Newton's mechanics. A new kind of spacetime symmtry arises.
Ordinary quantum mechanics is indifferent to the choice
of a potential, say in Schr\"odinger equation, as well.
However in physics of elementary particles where 
{\it a natural type of sub-atomic forces act},  we can 
not be sure that equations and symmetry of the ordinary quantum theory may be concerved. In particular, ordinary quantum theory points out that general relativity (GR) is 
negligible for spatial distances up to the Planck scale $l_P=(\hbar G/c^3)^{1/2} \sim 10^{-33} cm$. But 
consistency in the foundations of the quantum theory requires a``soft'' spacetime structure of the GR at essentially longer length. However, for some reasons 
this appears to 
be not enough. A new framework (``superrelativity'') 
for the desirable generalization of the foundation of quantum theory has been proposed 
\cite{Le1,Le2,Le3,Le4,Le5}.

The equivalence principle of Einstein is based on the
experimental fact that acceleration of bodies in the
gravitation field is independent on masses of these
bodies \cite{Ein}. This situation is physically 
equivalent
to the motion of the system of bodies in accelerated 
frame. We can not, of course, put the {\it criterion of  
identical acceleration} in the basis of geometrization
of quantum physical fields.  In quantum field theory the
notion of ``acceleration'' is poor at best and ambiguous
at worst because quantum particles have an internal
structure. Furthermore, at a deeper level there is
{\it entanglement  and even indistinguishability of
``internal'' and ``external'' degrees of freedom}.
Therefore in quantum regime we can not act 
{\it literally} as Einstein in GR but only in his 
{\it spirit} \cite{Le1,Le2,Le3,Le4,Le5}.
  
The notions of {\it material point, event, and classical
spacetime} in both special relativity (SR) and GR are 
liable to lead to confusion at the quantum level. 
Insead of these objects we use a set of 
new primordial elements. Namely, they are {\it pure 
quantum state, quantum transition, and quantum state
space}, respectively.  These main elements provide
a basis for a generalization of the equivalence principle
in quantum area. The principle of superrelativity 
states as follows: 
{\bf The general unitary motion of pure 
quantum states may be locally reduced to the geodesic 
motion in the projective Hilbert space by the 
introduction of a gauge (compensation) field which 
arises from the local change of a functional frame in 
the original Hilbert state space}.

We have put at the basis of our theory the fact that 
in {\it all interactions of quantum (``elementary'') 
particles  there is a conservation law of the 
electric charge}. Then unitary group SU(N+1) takes 
the place of {\it dynamical
group} because in the general case $g|\Psi> \neq |\Psi>$
and, of course, $g^{-1}Hg \neq H$ \cite{Feng}.  
There is the group of isotropy $H=U(1)_{el} \times U(N)$
of a fixed pure quantum state $e^{i\gamma}|\Psi>$.
That is transformations which effectively act on this ray lie in the coset $G/H=SU(N+1)/S[U(1)\times U(N)]$.
It is clear that deformations of the pure 
quantum state are due to some physical interaction;
the effect of the interaction has the geometric 
structure of a coset \cite{Le1}, i.e. the structure of 
the CP(N): $G/H=SU(N+1)/S[U(1)_{el} \times U(N)]=CP(N)$ \cite{KN}. This paves the way to the invariant
study of the spontaneously broken unitary symmetry.
This statement has a general character and does not 
depend on particular properties of the pure quantum
state. The reason  for the change of motion of material
point is an existence of a force. The reason for the 
change of a pure quantum state is an interaction which 
may be modeled by unitary transformations from the coset
G/H. The reaction of a material point is an acceleration. The reaction of a pure quantum state is its deformation
whose geometry should be discussed now 
\cite{Le1,Le2,Le5}.  

\section{Break-down of the projective symmetry}
Unitary symmetry in quantum mechanics is
expanded to the projective symmetry due to arbitrary normalization of state vector \cite{Feng}. 
But some interactions, gravity, for example, ban this 
degree of freedom and restore (with a reconstruction) 
the unitary symmetry. Furthermore, the global unitary
summetry is broken down to the isotropy subgroup
whose elements leave of some fixed ray of state
$e^{i\gamma}|\Psi>$ intact. In order to uderstand
a new type of symmetry arising here, 
we will discuss the infinitesimal 
interval in Hilbert space $\cal H\rm=C(N+1)$ which 
related to some linear Hermitian traceless operator 
$\hat{D} \in AlgSU(N+1)$. This operator creates some interval
$d_{\hat{D}}l^2=d\Psi_a^* d\Psi^a=<d\Psi|d\Psi>=$
$<\Psi|\hat{D}^+ \hat{D}|\Psi>d\epsilon^2$,
where $\epsilon$ is real-number group parameter,
and $\Psi^a$ are Fourier components of a decomposition 
of a state vector in some orthogonal basis
$|\Psi>=\sum_{a=0}^N \Psi^a|a>$ where 
$\sum_{a=0}^N |\Psi^a|^2=R^2$.
Now one can reduce the quadric to the 
principle axes by ansatz of ``squeezing'' full state 
covector to the vacuum form as in \cite{Le4}
\begin{equation}
<\Psi_0|=R(\Psi) e^{i\omega(\Psi)}(1,0,...,0).
\label{vac} 
\end{equation}
That is $|\Psi_0>=\hat{G}^{-1}|\Psi>$, where 
$\hat{G}=\hat{G}_1 \hat{G}_2...\hat{G}_N$ are matrices
which described in \cite{Le4}.  Since in the process 
of a ``squeezing'' gauge transformations from isotropy 
group should be 
accompanied by transformations from the coset $SU(N+1)/S[U(1)\times U(N)]$, in contrast to the 
gauge transformations of Ref. \cite{Gold} one has 
an {\it observable deformation of the quantum state}.
The gauge field is a physical evidence of this 
deformation due to the non-trivial topology of CP(N). 
Then one has
\begin{equation}
d_{\hat{D}}l^2=<\Psi_0|\hat{D'}(\Psi)^+ \hat{D'}(\Psi)|\Psi_0>=
R^2(|D'_{00}(\Psi)|^2+\sum_{i=1}^N |D'_{0i}(\Psi)|^2)d\theta^2.
\label{quad}
\end{equation}
The first term is interval along
the subalgebra of the isotropy group of the vector
$|\Psi_0>$ and the second one is interval in the tangent space to the coset itself. One-parameter tranformation
from the coset is a geodesic flow.  For the covector (\ref{vac}) the geodesic flow is generated by the 
operator 
\begin{eqnarray}
\hat{B}= \left(
\matrix{
0&f^{1*}&f^{2*}&.&.&.&f^{N-1*} \cr
f^1&0&0&.&.&.&0 \cr
f^2&0&0&.&.&.&0 \cr
.&.&.&. &.&.&. \cr
.&.&.&. &.&.&. \cr
.&.&.&. &.&.&. \cr
f^{N-1}&0&0&.&.&.&0
}
\right ),
\label{B} 
\end{eqnarray}
where $f^i=D'_{0i}(\Psi)$.
The flow is given by the unitary matrix
$\hat{T}(\tau,g)=\exp(i\tau\hat{B})$
where $g=\sqrt{\sum_{k=1}^{N}|f^k|^2},\Theta=g\tau$
\cite{Le1}.
Note, elements of tangent space to the coset will be functions of state vector  during the process of the ``squeezing'' . In that sense local (in functional space) dynamical variables arise. But invariant properties of 
the interval should be independent from a particular 
choice of the dynamical variable $\hat{D}$. The infinitezimal interval $\delta L^2$ in local projective coordinates may exemplify this independece. 
The generalized stereographic projection from the center
of the sphere $\sum_{a=0}^N |\Psi^a|^2=R^2$ onto the 
complex hyperplane $\Pi$ give us relationships between
coordinates of a point of the sphere in the original 
Hilbert space $\Psi^0,...,\Psi^a,...,\Psi^N$ and 
coordinates
$\Pi^1,...,\Pi^i,...,\Pi^N$
of its projection onto the hyperplane
\begin{equation}
\Psi^0=\lambda(R,\Pi)R,\quad \Psi^1=\lambda(R,\Pi)\Pi^1,
\Psi^2=\lambda(R,\Pi)\Pi^2,...,\Psi^N=\lambda(R,\Pi) \Pi^N
\label{psi}
\end{equation}
Then one has mapping $f:C(N+1) \to CP(N)$
for ($1 \leq i \leq N$)
\begin{equation}
\Pi^1=R\Psi^1 / \Psi^0,
\Pi^2=R\Psi^2 / \Psi^0,...,
\Pi^i=R\Psi^i / \Psi^0,...,
\Pi^N=R\Psi^N / \Psi^0
\label{pi}
\end{equation}
and
$\lambda^2 (\sum_{i=1}^N |\Pi^i|^2+R^2)=R^2$.
The ``squeezing factor'' $\lambda(R,\Pi)$ one can 
express from this equation
\begin{equation}
\lambda(R,\Pi)=R (\sum_{s=1}^N |\Pi^s|^2+R^2)^{-1/2}.
\label{lambda}
\end{equation}
Hereafter we will use only finite value of indexes 
$0 \leq a,b,...,d \leq N$ and $1 \leq i,k,...,s \leq N $,
remembering that in typical cases the limit 
$N \to \infty $ may be studied.
Now we can express homogeneous coordinates $\Psi$ in 
local coordinates $\Pi$:
\begin{equation}
\Psi^0=R^2 (\sum_{s=1}^N |\Pi^s|^2+R^2)^{-1/2},...,
\quad 
\Psi^i=\Pi^i R (\sum_{s=1}^N |\Pi^s|^2+R^2)^{-1/2}.
\label{psi1}
\end{equation}
It is easy to evaluate ($a=0$)
\begin{equation}
\frac{\partial \Psi^0}{\partial \Pi^i}=-\frac{1}{2}
R^2 \Pi^{*i} \left(\sqrt{\sum_{s=1}^N |\Pi^s|^2+R^2}\right)^{-3}, c.c.
\label{grad}
\end{equation}
and for other components ($a \geq 1$) one has
\begin{eqnarray}
\frac{\partial \Psi^a}{\partial \Pi^i}     & = &
R\left(\delta^a_i(\sum_{s=1}^N |\Pi^s|^2+R^2)^{-1/2}-
\frac{1}{2} \Pi^a \Pi^{*i}
(\sum_{s=1}^N |\Pi^s|^2+R^2)^{-3/2}\right),
c.c.
\label{partial}
\end{eqnarray}
Therefore one can express infinitezimal invariant interval
in the original Hilbert space as followes
\begin{equation}
\delta L^2= \delta_{ab}\delta \Psi^a \delta \Psi^{*b}= G_{ik*}\delta \Pi^i \delta \Pi^{*k}=\sum_a \frac{\partial \Psi^a}{\partial \Pi^i}
\frac{\partial \Psi^{*a}}{\partial \Pi^{*k}} 
\delta \Pi^i \delta \Pi^{*k}.
\label{interval}
\end{equation}
That is the generalized metric tensor of the original flat
Hilbert space in the local coordinates $\Pi$ is 
\begin{equation}
G_{ik*}=\sum_{a=0}^N \frac{\partial \Psi^a}{\partial \Pi^i}
\frac{\partial \Psi^{*a}}{\partial \Pi^{*k}}=
R^2 \frac{(\sum_{s=1}^N |\Pi^s|^2+R^2)\delta_{ik}-
\frac{3}{4}\Pi^{*i}\Pi^k}{(\sum_{s=1}^N |\Pi^s|^2+R^2)^2}.
\label{metric}
\end{equation}
For small $\Pi^i$, i.e. under small deformations of the
initial state $|\Psi_0>$ in a physical experiment
the full invariant interval
$\delta L^2$ in original Hilbert space is  numerically 
very close to the interval
\begin{equation}
dl^2=R^2\frac{(\sum_{s=1}^N |\Pi^s|^2+R^2)\delta_{ik}-
\Pi^{*i}\Pi^k}{(\sum_{s=1}^N |\Pi^s|^2+R^2)^2}
\delta \Pi^i \delta \Pi^{*k}
\label{distance}
\end{equation}
in the projective Hilbert space CP(N). Notwithstanding
I think that it is possible to find a trace of the 
non-zero sectional curvature of CP(N) in optic 
experiments for the measuring of Aharonov-Anandan
phase for light rays with different parameters. 
It will be discussed elsewhere.

The complexified Poisson bracket of the Fourier 
components in the Hilbert space relative to 
the local coordinates $(\Pi,\Pi^*)$ 
\begin{equation}
\{\Psi^a,\Psi^b\}=\frac{\partial \Psi^a}{\partial \Pi^i}\frac{\partial \Psi^{*b}}{\partial \Pi^{*i}}-\frac{\partial \Psi^{*a}}{\partial \Pi^{*i}}\frac{\partial \Psi^{b}}{\partial \Pi^{i}}
\label{poisson}
\end{equation}
creates the simplectic structure. For $a,b \geq 1$ one has:
\begin{equation}
\{\Psi^i,\Psi^k\}= 
\frac{R^2}{(\sum_{s=1}^N |\Pi^s|^2+R^2)^2}
\left[1-\frac{1}{4}\frac{\sum_{s=1}^N |\Pi^s|^2}{\sum_{s=1}^N |\Pi^s|^2+R^2}\right]
(\Pi^{*i}\Pi^k-\Pi^{*k}\Pi^i).
\label{bracket}
\end{equation}
Here we assumed that R has physical meaning and may be observable. This situation should be realized presumably 
in quantum field theory and theory of elementary 
particles where gravity should be taken into account and 
the superposition principle does not act \cite{Le1,Le2,Le3,Le4,Le5}.
Then the projective symmetry is broken. The unitary symmetry is local and hidden. {\bf In this case naturaly think that $\Psi^a$ describe not amplitudes of probability in some ensemble, but they should be Fourier amplitudes of the single scalar field carrier of energy-momentum and charge degrees of freedom}. 

\begin{figure}[ht]
\centerline{\epsffile{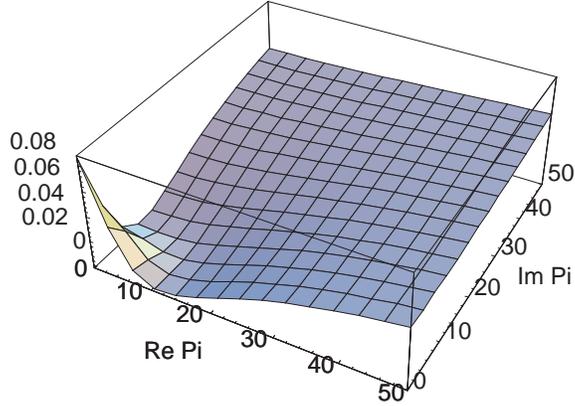}}
\caption{The connection (2.15) in the case 
CP(1). The value of $1/R+\Gamma^i_{kl}$ is shown. 
Radial directions are related to the Goldstone modes of ``deformations'' of quantum state. Angular rotations represent the Higgs modes of the gauge transformations 
in every point of a geodesic. All geodesics are subjected 
to a ``rigid rotation'' under the action of the isotropy 
group.}
\label{fig.1}
\end{figure}

Therefore it would be incorrect to interpret 
{\it literally} the formal existence of the generalized 
Hamilton structure as an evidence of spacetime motion.
The second quantization scheem should be 
changed as well (it will be discussed soon). The 
question is: what is the relationships between this 
``hidden dynamics'' in CP(N) and dynamics in ordinary spacetime? 
It may be shown that the natural connection in CP(N) 
\begin{equation}
\Gamma^i_{kl} = -2 (\delta^i_k \Pi^{l*} + \delta^i_l \Pi^{k*})(R^2 + \sum_s^{N} |\Pi^s|^2)^{-1}
\label{connection} 
\end{equation}
which is corresponds to the Fubini-Study metric (\ref{metric})
plays an important role in gauge transformations of
the functional frame. Namely, we will show that {\bf the connection (\ref{connection}) determines 
the natural intrinsic gauge potential of a local frame rotation in a tangent space of CP(N) and, therefore, modification of field dynamical variables. Relationships between the Goldsone and Higgs modes arise in an 
absolutely natural way} (see Fig.\ref{fig.1}).
\vskip .2cm
\vskip .2cm
Below we shall discuss a simple model of a ``wrapping'' of this underlying structure into the gauge (compensation) field in the reference Minkowski spacetime.

\section{Generalized nonlinear Klein-Gordon equation}
{\it The key idea is associated with a model 
for a single quantum particle as an extended field 
dynamical system. We will to interpret internal local coordinates (\ref{pi}) as relative Fourier componets 
of some wave packet in reference Minkowski spacetime. Therefore the ``Hamilton pre-system'' in CP(N) is ``wrapped'' in the shell of a gauge field in ordinary spacetime and one has, hence, a non-local extended 
dynamical field configuration}.
It should play the role of a model 
of non-local quantum particles in the framework of the causal approach to quantum theory \cite{Le1,Le2}.
For the global `shaping'' of this configuration, 
the Klein-Gordon field configuration with ``Poincar\'e-radial'' Lagrangian for non-linear interaction has been chosen. This Lagrangian depends only on the ``radial'' variable $\rho$ , i.e.
$\Phi = \Phi (\rho)$, where $\rho^2 =  x_\mu x^\mu$ and $x^\mu$ 
Since 
$\Phi_{,\mu} = \frac{\partial \Phi}{ \partial x^ \mu}=
\frac {x_ \mu}{\rho} \frac{d \Phi}{d \rho}$ and
$\Phi^{,\mu} = \frac {\partial \Phi}{ \partial x_ \mu}=
\frac {x^ \mu}{\rho} \frac{d \Phi}{d \rho} $, a Lagrangian density may be written as
\begin{equation}
\cal L\rm=\frac{1}{2} \Phi^*_\mu \Phi^\mu - U(\Phi(\rho)) = 
\frac{1}{2} \left|\frac{d\Phi}{d\rho} \right|^2 -  U(\Phi(\rho)),
\label{lagr}
\end{equation}
where we have assumed a general form for the effective self-interaction
term $U(\Phi(\rho))$. The equation of motion of the scalar field
acquires the form of the ordinary differential (nonlinear in general)  equation
\begin{equation}
\frac{d^2 \Phi^*}{d \rho^2} + (3/\rho) \frac{ d \Phi^*}{d \rho} +
2\frac{\partial U(\Phi(\rho))}{\partial \Phi} =0. 
\label{eqgen}
\end{equation}
This equation has been
used in the problem of vacuum decay (see (\cite{W2})
and references therein). If the potential $U(\Phi(\rho))$ has the form 
$U(\Phi(\rho))=
\frac{1}{2}(mc/\hbar)^2\Phi^* \Phi =\frac{1}{2}\alpha^2 |\Phi|^2$,
then (\ref{eqgen}) is the linear Lommel equation 
\begin{equation}
\frac{d^2 \Phi^*}{d \rho^2} + (3/\rho) \frac{ d \Phi^*}{d \rho} + \alpha^2 \Phi^* = 0, 
\label{lommel}
\end{equation}
for which a solution is expressed in the Bessel function 
$\Phi = \rho^{-1} J_{-1} (\alpha \rho)$.
{\bf However  a ``deformation'' of  these solutions into solutions of some effectively nonlinear Klein-Gordon equation by the geodesic flow is interesting
for our purpose}. If we choose the the classical radius of the electron $r_0=\frac{e^2}{mc^2}$
as the unit of distance scale $\rho = x r_0$, 
then $(mc/\hbar)^2$ in (\ref{lommel}) becomes the fine structure constant 
$\alpha = \frac{e^2}{\hbar c}$.  Let's suppose $y=(\rho/r_0)^2$.
One can represent a solution of (\ref{lommel}) in the 
$y$-variable as a Fourier series
\begin{equation}
|\Phi> = x^{-1} J_{-1} (\alpha x)=
 \sum_{k=0}^{\infty} \Phi^k |k,y>,
\label{solution}
\end{equation}
where 
\begin{equation}
\Phi^m=\frac{-1}{2^m m! \sqrt\pi}\int_{-\infty}^{\infty}
dy \exp(-\frac{y^2}{2}) H_m(y)y^{-1/2} J_1(\alpha y^{1/2})
\label{fourier}
\end{equation}
and
$|n,y>=(2^n n! \sqrt\pi)^{-1/2}\exp(-y^2 /2)
H_n(y)$
is complete set of orthogonal Hermitian functions 
on the interval $(-\infty,\infty)$.
Our geodesic flow acts on these Fourier components. It is convenient to transform the state covector (\ref{fourier})
to the ``vacuum'' form (\ref{vac})
by the set of matrices $\hat{G}$ as followes \cite{Le4}.
In order to find the elements $f^i$ of
the generator of the geodesic flow for deformation of
the vacuum state toward the solution (\ref{solution})
one should note that if result of the periodic geodesic ``deformation'' of the  initial solution of 
\begin{equation}
|\Psi (\tau,g,y)> = \sum^{\infty}_{m,n=0} |n,y> \Phi^m [\hat{G} \hat{T}(\tau,g) 
\hat{G}^{-1}]^n_m
\label{droplet}
\end{equation}
is not so far from (\ref{vac}), then one can span them 
by an unique geodesic 
$(1,0,...,0)\hat{T}(\tau,g)=R^{-1}(\Phi^0,\Phi^1,...,\Phi^N)$.
It may be shown that $cos\Theta=|\Phi^0|/R$, 
$|f^i|=g|\Phi^i|(R^2-|\Phi^0|^2)^{-1/2}$ and 
$\arg f^i=\arg \Phi^i$ (up to the general phase). Thus
``rate'' and direction of the transformation of the 
vacuum vector into the solution of the Lommel equation is 
determined by the matrix 
$\hat{P}=\hat{G}^{-1}(\Phi)\hat{B}(\Phi)\hat{G}(\Phi)$.

In order to establish relationships  between ``internal'' parameters in $CP(\infty)$ and a propagation of the 
scalar field near the light cone in the 
``reference spacetime'', we should ``lift'' a geodesic deformation of the initial Fourier components into the 
fiber bundle. Namely, if we assume that {\it in accordance with the ``superequivalence principle'' an infinitesimal geodesic ``shift'' of field dynamical variables could be compensated by an infinitesimal transformations of the 
basis in Hilbert space, then one can get some effective self-interaction potential as an addition to the mass 
term in original Klein-Gordon equation in the Lommel's 
form} (\ref{lommel}).  
We will label hereafter vectors of the Hilbert space by Dirac's notations
$|...>$ and tangent vectors to CP(N) or  $CP(\infty)$ 
(field dynamical variables) by 
arrows over letters, $\vec \xi \in T_{\Pi}CP(\infty)$, for example. Then one has a definition
of the rate of a state vector  changing
$|v(x)>=-(i/\hbar)\hat{P}|\Psi (x)>$.
The ``descent'' of the vector field $|v(x)>$ onto the base manifold 
$CP(\infty)$ is a mapping by the two formulas:
$f:\cal H \rm \to CP(\infty)$, i.e. like (\ref{pi})
$f:(\Psi^0,..., \Psi^a,...)=(R\frac{\Psi^1}{\Psi^0},..., R\frac{\Psi^i}{\Psi^0},...)=(\Pi^1,...,\Pi^i,...)$
and
\begin{eqnarray}
\vec \xi=f_{*(\Psi^0,..., \Psi^m,...)} |v(x)> 
=\frac{d}{d \tau}(R \frac{\Psi^1}{\Psi^0},..., R \frac{\Psi^i}{\Psi^0},...)\Bigl|_0 \cr
=-(i/\hbar)[P^1_0+(P^1_k-P^0_k \Pi^1)\Pi^k,..., P^i_0+(P^i_k-P^0_k \Pi^i)\Pi^k,...].  
\label{map2}
\end{eqnarray}
This operator determines a field dynamicl variable
$\vec \xi$ (\ref{map2}). At a point  
$\Pi+\delta \Pi$ in $CP(\infty)$ the ``shifted'' field  
$\vec \xi+ \delta \vec \xi$
$=\vec \xi+\frac{\delta \vec \xi}{\delta l} \delta l$
contains the derivative $\frac{\delta \vec \xi}{\delta l}$, which is not, in the
general case, a tangent vector to $CP(\infty)$, but the 
{\it covariant derivative}
$\frac{\Delta \xi^i}{\delta l}$
$=\frac{\delta \xi^i}{\delta l}+\Gamma^i_{km}
\xi^k \frac{\delta \Pi^m}{\delta l}$
is a tangent vector to $CP(\infty)$. Now we should ``lift'' the new tangent
vector $\xi^i + \Delta \xi^i$ into the original Hilbert space $\cal H \rm$,
that is, one needs to realize two inverse mappings: 
$f^{-1}:CP(\infty) \to \cal H \rm $
\begin{eqnarray}
f^{-1}(\Pi^1+\Delta \Pi^1,...,\Pi^i + \Delta \Pi^i,...) 
=[\Psi^0,\Psi^1+\Psi^0 \Delta \Pi^1,...,\Psi^i+\Psi^0 
\Delta \Pi^i,...]
\label{map-1}
\end{eqnarray}
and then
\begin{eqnarray}
f^{-1}_{* \Pi+\delta \Pi}(\vec \xi + \Delta \vec \xi)
=[v^0, v^1+ \frac{d}{d\tau}(\Psi^0 \Delta \Pi^1)
\Big|_0,...,  v^i+\frac{d}{d\tau}(\Psi^0 
\Delta \Pi^i)\Big|_0,...].
\label{map-2}
\end{eqnarray}
It may be shown in our original Hilbert space $\cal H \rm$ 
that the term $|dv>$ arises as an additional rate of a change of some general state vector $|\Psi>$
\begin{eqnarray}
|dv>=-\frac{i}{\hbar}d\hat{P}|\Psi> 
= [0, \frac{d}{d\tau}(\Psi^0 \Delta \Pi^1)\Bigl|_0|1,x>,...,\frac{d}{d\tau}(\Psi^0 \Delta \Pi^i)\Bigl|_0|i,x>,...],
\label{dv} 
\end{eqnarray}
where $\Delta \Pi^i=-\Gamma^i_{km}\xi^k d\Pi^m \tau$.
Then $\Delta U=\frac{\delta U}{\delta \Pi^i}d \Pi^i
+\frac{\delta U}{\delta \Pi^{*i}}d \Pi^{*i}$
where $\frac{\delta U}{\delta \Pi^m}=-\hbar \Gamma^i_{km}\xi^k |i,y>$ may be  treated as an ``instantaneous'' 
self-interacting potential of the scalar configuration associated with 
the infinitesimal gauge transformation of the local frame  
with coefficients (\ref{connection}).
That is the state vector (\ref{map-1}) inherits the geomeric structure
of the $CP(\infty)$ and perturbed Lagrangian is as follows:
\begin{eqnarray}
\cal L'\rm=\cal L\rm(\Psi+\Delta \Psi)=(\Psi+\Delta \Psi)^{m*}\frac{\partial <m,y|}{\partial t}
\frac{\partial |n,y>}{\partial t}(\Psi+\Delta \Psi)^{n}\cr
-(\Psi+\Delta \Psi)^{m*}\nabla <m,y|\nabla |n,y>(\Psi+\Delta \Psi)^n \cr
-\alpha^2 (\Psi+\Delta \Psi)^{m*}<m,y|n,y>(\Psi+\Delta \Psi)^n,
\label{newlagr}
\end{eqnarray}
where $\Delta \Psi^i=- \Psi^0 \Gamma^i_{km}\xi^k d\Pi^m \tau$ \cite{Le1}.
Note, spacetime derivatives  of the basis Hermitian functions 
$\nabla|n,y>$, $\frac{\partial |n,y>}{\partial t}$
only arise in the formula for Lagrangian density. On the 
other hand, only Fourier components (\ref{fourier})  are subjected to the variation by the geodesic flow.
Equation of motion of the self-interacting field configuration-droplet may be obtained from variation
of the Lagrangian relative to the variation of
$|\Psi>$. One then has a generalized Klein-Gordon
equation
\begin{eqnarray}
\Box \Psi^*+ \Box A^* +\Psi^*_{\mu \mu}\frac{\delta A_{\mu}}
{\delta \Psi_{\mu}}+\Psi^*_{\mu}\frac{\delta A_{\mu \mu}}
{\delta \Psi_{\mu}} +\alpha^2(\Psi^*+\Delta\Psi^*+
\Psi^*\frac{\delta\Delta\Psi}{\delta \Psi})=0, c.c.,
\label{nlkg}
\end{eqnarray}
where $A_{\mu}=\frac{\partial \Delta \Psi}
{\partial x^{\mu}}$,$A_{\mu \mu}=\frac{\partial \Delta \Psi}
{\partial x^{\mu} \partial x^{\mu}}$, $\Psi_{\mu}=\frac{\partial \Psi}
{\partial x^{\mu}}$,$\Psi_{\mu \mu}=\frac{\partial \Psi}
{\partial x^{\mu} \partial x^{\mu}}$, and
for enough small $\tau$ 
\begin{equation}
\Delta \Psi^i=- g\Psi^0\tau^2
(1+\frac{|\Psi^0|^2}{R^2})^{-1/2}\Gamma^i_{km}\xi^k \Psi^m.
\label{delta}
\end{equation}
If the radius $R$ of the sectional curvature $1/R^2$ of the projective Hilbert space goes to infinity, one obtains the ordinary Klein-Gordon equation.
We can hope that non-trivial metric and topology of the projective Hilbert space endows global solutions of non-linear wave equation with interesting physical properties.
In the general case the curvature of the 
projective state space influences the wave dispersion 
of a nonliner solution of the equation. It may be
treated as a base of the experimental testing of a 
quantum nonlinearity. This topic will be investigated 
in the near futures.

ACKNOWLEDGEMENTS

I thank Yuval Ne'eman and Larry Horwitz  for numerous useful
discussions and Yakir Aharonov for attention to this work.
This research was supported in part by grant PHY-9307708
of the National Science Foundation, and by grant of the
Ministry of Absorption of Israel.

\end{document}